\documentclass[a4paper,11pt]{article}
\usepackage[utf8x]{inputenc}
\usepackage{bm}
\usepackage{amsthm,amsmath,amssymb,amsfonts,amssymb,hyperref} 
\usepackage{booktabs}
\usepackage{enumerate}
\usepackage{subfigure}
\usepackage{graphicx}
\usepackage{enumitem}

\usepackage{lmodern}
\usepackage[left=1in,right=1in,top=1.5in,bottom=1.5in]{geometry}
\usepackage{babel}
\usepackage{setspace}
\usepackage{natbib}
\usepackage[draft]{fixme}
\usepackage{mathtools}
\usepackage{hyphenat}
\usepackage[section]{placeins}

\DeclareMathOperator{\E}{\mathbb{E}}    % Expected value
\newcommand{\M}{\bm}
\newcommand{\V}{\bm}

\usepackage{url}



\usepackage{dcolumn}
\newcolumntype{e}{D{.}{.}{9}}

\numberwithin{equation}{section}

\theoremstyle{plain}

\theoremstyle{definition}

\theoremstyle{remark}

\title{Portfolio Allocation under Asymmetric Dependence in Asset Returns using Local Gaussian Correlations}
\author{Anders D. Sleire$^2$, B\aa rd St\o ve$^2$\thanks{Corresponding author. Tel. +47 55 58 28 86. E-mail: Bard.Stove@math.uib.no.}, H{\aa}kon Otneim$^1$, Geir Drage Berentsen$^1$, \\ Dag Tjøstheim$^2$, Sverre Hauso Haugen$^2,$  \\ \\
$^1$Department of Business and Management Science, Norwegian School of Economics\\
$^2$ Department of Mathematics, University of Bergen, Norway
}

\begin{document}
     
\maketitle

\begin{abstract}

It is well known that there are asymmetric dependence structures between financial returns. In this paper we use a new nonparametric measure of local dependence, the local Gaussian correlation, to improve portfolio allocation. We extend the classical mean-variance framework, and show that the portfolio optimization is straightforward using our new approach, only relying on a tuning parameter (the bandwidth). The new method is shown to outperform the equally weighted ("1/N") portfolio and the classical Markowitz portfolio for monthly asset returns data.

\end{abstract}

\section{Introduction}

In modern portfolio theory the aim is to allocate assets by maximizing the expected return of the porfolio while minimizing its risk, for instance measured by the standard deviation. \cite{markowitz1952} provides the foundation for this mean-variance (MV) approach, where the crucial assumption is that the asset returns follow a joint-Gaussian distribution, and hence that the dependence between these returns is fully described by the linear correlation coefficient. The idea is simple; low correlated assets are suitable for diversification, while highly correlated assets should be avoided.  

However, as shown by numerous empirical studies, it is well documented that there are asymmetries in the distribution of
financial returns. This is especially true when the market is going down, which often leads to stronger dependence between assets, known as asymmetric dependence structures, see e.g. 
\citep{Silvapulle2001, Campbell2002, Okimoto2008,Ang2002b,Hong2007,Chollete2009,Garcia2011}. Other forms of asymmetries are also present; asymmetric volatility clustering, see e.g. \cite{glosten1993} or skewness within the distribution of stock returns, see e.g. \cite{patton2004} or \cite{ait2001}.
The asymmetric dependence implies that the classical mean-variance optimized portfolios are not efficient with respect to their effective risk profile. Thus the benefit of diversification will erode if the correlations are asymmetric.  Several studies have sought to overcome this shortcoming by modeling the dependence structure by using copula theory, and employing it in the optimization of a portfolio. Among the first papers introducing such an idea, \cite{patton2004},  examines whether the asymmetries are predictable and if portfolio decisions are improved by forecasting such asymmetries. He finds that for investors with no short-sales constraints, the knowledge of these asymmetries leads to economic gains. 
\cite{hatherley2007} find that managing asymmetric dependence, using a Clayton copula against the multivariate Gaussian model, reduces downside exposure. 
 \cite{low2013} make use of the bivariate Clayton and the
Clayton canonical vine copulas to address the asset allocation for
loss-averse investors through the minimization of CVaR in portfolios of up to 12 constituents. \cite{kakouris2014} employ a mixture of copulas to derive CVaR and the worst case CVaR used to optimize a convex portfolio of stock indices. \cite{Bekiros2015} use pair-vine copula models and minimum risk optimal portfolios applied to mining stock portfolios. Finally, \cite{han2017} use a copula-GARCH and DCC copulas approach, and extend \cite{kakouris2014} to dynamic portfolio optimization models. 

However, these procedures are in most cases quite complicated, and there is no guarantee that portfolio allocations based on complex models will perform better (see e.g. \cite{demiguel2007} and \cite{low2016}, where they find that outperforming the naive $1/N$ portfolio remains an elusive task). A non-technical asset manager might be overwhelmed by such choices.

We propose a much simpler approach. Without making assumptions about the nature of the underlying probability model, we present an adjustment to the correlation matrix of the assets, that takes the current state of the market into account by using the local Gaussian correlation (LGC), which is a recently developed dependence measure capable of detecting non-linear and asymmetric dependence patterns, see \cite{Hufthammer2008a}. In other words, if the market is going down and correlations increase as is often loosely stated as a fact, this effect, being a consequence of the nonlinearity/asymmetry of the dependence between assets, will be captured by the LGC. It will thus provide an updated correlation matrix to be used in the portfolio allocation problem. Our proposed procedure relates to other studies that use a dynamic model for the correlation matrix to improve the asset allocation, see e.g. \cite{engle2006},  
\cite{kalotychou2014} and \cite{aslanidis2013}. But, our procedure differs, as the main goal of this paper is to extend the classical MV framework by using the theory of local Gaussian correlation. % introduced in \cite{Hufthammer2008a}. 

The organisation of the paper is as follows. In section 2, we briefly review the local Gaussian correlation, while section 3  present the classical mean-variance portfolio approach and the extension using the local Gaussian correlation. 
In section 4 we analyse a data set consisting of several asset returns, and investigate the performance of portfolios constructed by our new approach and other methods. Finally, section 5 offers some conclusions and outlook.

\section{Local Gaussian correlation} \label{LGC} 

This paper relies on the recent developed dependence measure, the local Gaussian correlation (LGC), introduced by 
\cite{Hufthammer2008a}. This is a local characterization of dependence in space, and this idea has also been extended to several different situations, as test of independence, see \cite{bere:tjos:2014}, \cite{laca:tjos:2017} and \cite{laca:tjos:2018}, density and conditional density estimation, see \cite{otne:tjos:2017} and \cite{otne:tjos:2018}, a local Gaussian partial correlation, \cite{otneim2021}, and local Gaussian spectral estimation, see \cite{jordanger2020}. Finally, the relationship between the local Gaussian correlation and different copulas has been studied in \cite{berentsen2014recognizing}. For completeness, we briefly present the local Gaussian correlation in a standard fashion, and we note that this section closely follows the presentation of the LGC in \cite{tjos:otne:stov:2021}. 

Let $\bm R=(R_1,R_2)$
represent the return on two risky assets with density $f(\bm r)=f(r_1,r_2)$. Next, a description of how $f$ can be approximated locally in a
neighbourhood of each point $\bm r=(r_1,r_2)$ by a Gaussian bivariate density, \(\psi_{r_1,r_2}(\V v)\), where \( \V v = (v_1,v_2)\) are running variables, is given. Let \(\bm \mu(\bm r) = (\mu_1(\bm r),\mu_2(\bm r))\) be the mean vector in the normal distribution having density \(\psi_{\bm r}\), \(\bm\sigma(\bm r) = (\sigma_1(\bm r),\sigma_2(\bm r))\) is the vector of standard deviations, and \(\rho(\bm r)\) is the correlation coefficient in the normal distribution \(\psi_{\bm r}\). The approximating density is then given as
\begin{align}
&\psi_{\bm r} = \psi(\bm v,\mu_1(\bm r),\mu_2(\bm r),\sigma_1^2(\bm r),\sigma_2^2(\bm r),\rho(\bm r)) = \frac{1}{2\pi\sigma_1(\bm r)\sigma_2(\bm r)\sqrt{1-\rho^2(\bm r)}} \nonumber \\
& \qquad\times \exp\Big[-\frac{1}{2} \frac{1}{1-\rho^2(\bm r)}\Big(\frac{(v_1-\mu_1(\bm r))^2}{\sigma_1^2(\bm r)}-2\rho(\bm r)\frac{(v_1-\mu_1(\bm r))(v_2-\mu_2(\bm r))}{\sigma_1(\bm r)\sigma_2(\bm r)} \nonumber \\ & \qquad\qquad  +\frac{(v_2-\mu_2(\bm r))^2}{\sigma_2^2(\bm r)}\Big) \Big]. 
\label{eq:bivariate-normal}
\end{align}
Moving to another point \(\bm r'=(r_1',r_2')\) gives another approximating normal distribution \(\psi_{\bm r'}\) depending on a new set of parameters \((\mu_1(\bm r'),\mu_2(\bm r'), \sigma_1(\bm r'), \sigma_2(\bm r'), \rho(\bm r'))\). One exception to this is the case where \(f\) itself is Gaussian with parameters \((\mu_1,\mu_2, \sigma_1, \sigma_2 ,\rho)\), in which case \((\mu_1(\bm r),\mu_2(\bm r), \sigma_1(\bm r), \sigma_2(\bm r), \rho(\bm r)) \equiv (\mu_1, \mu_2, \sigma_1, \sigma_2, \rho)\).

The population parameter vector, \(\bm \theta(\bm r) \stackrel{\textrm{def}}{=} (\mu_1(\bm r),\mu_2(\bm r),\sigma_1(\bm r),\sigma_2(\bm r), \rho(\bm r))\), are obtained by minimizing the local penalty function 
 measuring the difference between \(f\) and \(\psi_{\bm r}\). It is defined by
\begin{equation}
q = \int K_{\V b}(\V v-\V r)[\psi(\V v,\V \theta(\V r))-\ln \{\psi (\V v,\V \theta(\V r))\}f(\V v)] \textrm{d} \V v
\label{eq:Kullback-distance}
\end{equation}
where \(K_{\V b}(\V v-\V x) = (b_1b_2)^{-1}K_1(b_1^{-1}(v_1-r_1))K_2(b_2^{-1}(v_2-r_2))\) is a product kernel with bandwidths $\V b = (b_1, b_2)$. As is seen in Hjort and Jones (1996, pp 1623-1624), the expression in \eqref{eq:Kullback-distance} can be interpreted as a locally weighted Kullback-Leibler distance from \(f\) to \(\psi(\cdot,\V \theta(\V r))\). Hence, the minimizer \(\V \theta_{\V b}(\V r)\) (also depending on \(K\)) should satisfy
\begin{equation}
\int K_{\V b}(\V v-\V r)\frac{\partial}{\partial \theta_j} [\ln\{\psi(\V v,\V \theta(\V r))\}f(\V v)-\psi(\V v,\V \theta(\V r))] \textrm{d} \V v = 0, \;\;j=1,\ldots,5.
\label{eq:score-equation}
\end{equation}\\
In the first step we define the population value \(\V \theta_{\V b}(\V r)\) as the minimizer of \eqref{eq:Kullback-distance}, assuming that there is a unique solution to \eqref{eq:score-equation}. The definition of \(\V \theta_{\V b}(\V r)\) and the assumption of uniqueness are essentially identical to those used in \cite{hjor:jone:1996} for more general parametric families of densities.

In the next step we let \(\V b \to \V 0\) and consider the limiting value \(\V \theta(\V r) = \lim_{\V b\rightarrow \V 0} \V \theta_{\V b}(\V r)\). This is in fact considered indirectly by \cite{hjor:jone:1996} and more directly in \cite{Hufthammer2008a}, both using Taylor expansion arguments. In the following we assume that a limiting value \(\V \theta(\V r)\) independent of \(\V b\) and \(K\) exists. 

In estimating \(\bm \theta(\bm r)\) and \(\bm \theta_{\bm b}(\bm r)\) a neighborhood with a finite bandwidth has to be used, this is in analogy with nonparametric density estimation. The estimate \(\widehat{\bm \theta}(\bm r) = \widehat{\bm \theta}_{\bm b}(\bm r)\) is then obtained from maximizing a local likelihood. Given observations \(\bm R_1,\ldots,\bm R_n\) the local log likelihood is determined by
\begin{align}
L(\bm R_1,\ldots,\bm R_n,\bm \theta(\bm r)) &= n^{-1}\sum_i K_{\bm b}(\bm R_i - \bm r)\log \psi(\bm R_i,\bm \theta(\bm r)) \nonumber \\ & \qquad\qquad\qquad - \int K_b(\bm v - \bm r)\psi(\bm v, \bm \theta(\bm r))\textrm{d} \bm v.
\label{eq:LGC-35}
\end{align}
When \(\bm b \to \infty\), the last term has 1 as its limiting value, and the likelihood reduces to the ordinary global likelihood. This last term is essential, as it implies that \(\psi(\bm r, \bm \theta_{\bm b}(\bm r))\) is not allowed to stray far away from \(f(\bm r)\) as \(\bm b \to \bm 0\). Indeed, using the notation
\begin{equation}
u_j(\cdot,\bm \theta) \stackrel{\textrm{def}}{=} \frac{\partial}{\partial \theta_j} \log \psi(\cdot,\bm \theta),
\label{eq:LGC-defu}
\end{equation}
by the law of large numbers, or by the ergodic theorem in the time series case, assuming \(\E(K_{\bm b}(\bm R_i - \bm r)\log \psi(\bm R_i, \bm \theta_{\bm b}(\bm r))) < \infty\), we have almost surely
\begin{align}
\frac{\partial L}{\partial \theta_j} &= n^{-1}\sum_i K_{\bm b}(\bm R_i - \bm r)u_j(\bm R_i, \bm \theta_{\bm b}(\bm r)) \nonumber \\ 
& \qquad - \int K_{\bm b}(\bm v - \bm r)u_j(\bm v,\bm \theta_{\bm b}(\bm r))\psi(\bm v,\bm \theta_{\bm b}(\bm r)) \textrm{d} \bm v \nonumber \\
&\to \int K_{\bm b}(\bm v - \bm r)u_j(\bm v,\bm \theta_{\bm b}(\bm r))[f(\bm v) - \psi(\bm v, \bm \theta_{\bm b}(\bm r))] \textrm{d} \bm v.
\label{eq:LGC-4}
\end{align}
Setting the expression in the first line of \eqref{eq:LGC-4} equal to zero yields the local maximum likelihood estimate \(\widehat{\bm \theta}_{\bm b}(\bm r)\) (\(=\widehat{\V\theta}(\bm r)\)) of the population value \(\bm \theta_{\bm b}(\bm r)\) (and \(\bm \theta(\bm r)\) which satisfies \eqref{eq:score-equation}).

An asymptotic theory has been developed in \cite{Hufthammer2008a} for \(\widehat{\V \theta}_{\V b}(\V r)\) for the case that \(\V b\) is fixed and for \(\widehat{\V \theta}(\V r)\) in the case that \(\V b \to \V 0\). The first case is much easier to treat than the second one. In fact for the first case the theory of \cite{hjor:jone:1996} can be taken over almost directly, although it is extended to the ergodic time series case in \cite{Hufthammer2008a}. In the case that \(\V b\rightarrow \V 0\), this leads to a slow convergence rate of \((n(b_1b_2)^{3})^{-1/2}\), which is the same convergence rate as for the the estimated dependence function treated in \cite{jone:1996}.

We have thus far concentrated on the bivariate case, in which we estimate a single local Gaussian correlation based on a bivariate sample. In principle, it is straightforward to extend to the case of more than two variables. Assume that we observe a multivariate sample $\V R_i = \{R_{1i}, \ldots, R_{pi}\}$, $i=1,\ldots, n$ with dimension $p>2$. We can then estimate the $p\times p$ local correlation matrix $\M \rho(\V r) = \{\rho_{k\ell}(\V r)\}$, $1\leq k< \ell \leq p$, $\V r = (r_1,\ldots,r_p)$, as well as the $p$ local means and local variances $\V \mu(\V r) = \{\mu_1(\V r), \ldots, \mu_p(\V r)\}$ and $\V \sigma(\V r) = \{\sigma_1(\V r), \ldots, \sigma_p(\V r)\}$ by maximizing the local likelihood function \eqref{eq:LGC-35}. The precision of such estimates, however, deteriorates quickly as the dimension $p$ grows, due to the curse of dimensionality. 

However, a simplifying technique that reduces the complexity of this estimation problem is to estimate each local correlation $\rho_{k\ell}(\V z)$ as a bivariate problem by only considering the corresponding pair of observation vectors $\{R_{ik}, R_{i\ell}\}$, $i=1,\ldots,n$. Thus, we reduce the $p$-variate problems of estimating the local parameters depending on all coordinates, to a series of bivariate problems of estimating pairwise local correlations depending on their respective pairs of coordinates. In this way, we obtain a simplification that is analogous to an additive approximation in nonparametric regression. This technique is applied in the empirical analysis that follows. For more details regarding this pairwise modeling approach, see \cite{otne:tjos:2017}. In particular, they show that the convergence speed is improved to $(nb^2)^{-1/2}$.

As already mentioned, the local estimates depend on the smoothing device - the bandwidth vector $\V b$ and a specific choice of the kernel function, $K$. In the empirical analysis, we use the Gaussian kernel, and the bandwidth selector used is the plug-in selector suggested in \cite{stov:tjos:huft:2014} — the global standard deviation of the observations times a constant equal to 1.1. 

Finally, we note that the local Gaussian correlation has been used in several studies examining the dependence structure between asset returns, and in testing for financial contagion, see e.g. \cite{stov:tjos:2014}, \cite{stov:tjos:huft:2014}, \cite{bampinas:2017} and \cite{nguyen:2020}.

\section{Portfolio allocation using local Gaussian correlation} \label{sec2} 

Mean-variance based portfolio construction is a common approach for asset management. Introduced by \cite{markowitz1952}, the measures of return and risk are the mean and variance of the portfolios' returns, respectively. Portfolios are considered mean-variance efficient if they minimize the variance for a given mean return or maximize the return for a given level of variance.  

 In this section we adopt the general formulation for portfolio optimization, which consists of minimization of a risk measure given a target reward and operational constraints. We assume there are $N$ risky assets. The returns on the risky assets are denoted by $\bm r_t \in \mathbb{R}^N$, which are assumed to have expected values $\bm \mu_t \in \mathbb{R}^N$, and covariance matrix $\M \Sigma_t \in \mathbb{R}^N\times\mathbb{R}^N$ of the portfolio of asset returns at time $t$. Further, let $\bm w_t \in \mathbb{R}^N$ be the unknown vector of optimal portfolio weights at time $t$.

The MV optimization problem is defined as follows; the weights of the chosen portfolio are given by a vector $\bm w_t$, invested in $N$ risky assets; and the investor selects $\bm w_t$ to maximize the expected quadratic utility function at each time $t$, that is
\begin{equation}\label{eq:5.7}
    \max_{\bm w_t} U = \bm w_t^T \bm \mu_t-\frac{\gamma}{2}\bm w_t^T\M \Sigma_t\bm w_t,
\end{equation}
where $U$ is the investor's utility, and $\gamma$ represents the investor's degree of risk aversion.  Hence, for a range of different risk aversion levels, the MV optimization will produce corresponding optimal portfolios with a trade-of between expected volatility and expected return. However, throughout this section, for simplicity, the risk aversion coefficient $\gamma$ will be fixed and equal to 1.

There are three different ways of formulating the MV optimization problem: minimize the risk subject to a lower bound on the expected return (which results in the Minimum Variance portfolio); maximize the expected return subject to an upper bound on the risk; optimize the corresponding ratio between risk and return subject to a given level of risk aversion. We assume that $w_1+\dots+w_N=1$. This constraint states that all capital must be invested in the portfolio (\emph{full investment} constraint), where the weights correspond to portions of the capital allocated to a given component. Another type of constraint is related to \emph{long only} positions, which specify that we can only buy shares and therefore only have position-related weights in contrast to the case of short positions, in which the selling positions would be reflected as negative weights. In the empirical example that follows, both cases will be examined. Further, we do not include a risk-free asset in our treatment of the portfolio allocation problem, but this will not impact our main findings.

The optimization problem with additional non-negativity constraints cannot be solved by the method of Lagrange multipliers because of the inequality constraints; it must be represented as a quadratic programming problem. Furthermore, the theory is unable to account for the presence of higher moments beyond the mean and variance in both the portfolio returns distributions or investor preferences, such as skewness and kurtosis. See e.g. \cite{francis:2013} or \cite{yao:2015} for a more detailed treatment of modern portfolio theory.

The typical portfolio allocation problem that arises in practice is described below. In the empirical analysis, we use monthly return data, but of course, shorter or longer time horizons are possible. As data of returns become available in time, we follow the same approach as \cite{demiguel2007}, \cite{tu:2011} and \cite{low2016} where rolling sampling windows of historical returns are used to estimate the expected return vector $\bm \mu_t$ and covariance matrix $\M \Sigma_t$ required as inputs into the Markowitz model. More specifically, the process is given as follows;

\begin{enumerate}
    \item At time $t$, a rolling sampling window of $M$ trading months is selected.
    \item During each month at time $t$, starting from $t=M+1$, the return data for the $M$ previous months are used to estimate the one month ahead expected return vector $\bm \mu_t$ and the covariance matrix $\M \Sigma_t$ by the standard empirical versions. As new information arrives at month $t+1$, these estimates are updated. This process is repeated by incorporating the return for each month going forward and ignoring the earliest one, until the end of the sample.
    \item Based upon these estimates, the various optimization problems are solved and the updated portfolio weights are updated at every first trading day of each month, and the rebalancing is done to construct a portfolio that achieves the desired investment objective.
    \item The estimates of $\bm w_t$ are then used to calculate out-of-sample returns $\hat{r}_{t+1}$ and portfolio performance over the next month. A total of $n-M$ out-of-sample returns are produced for each model, with $n$ being the total number of observations.
    \item These out-of-sample returns and portfolio weights are analyzed using a range of performance metrics and statistical measures that are reported for each model, respectively. For example, one can examine the cumulative returns resulting from a one dollar initial investment after a specified end date.
\end{enumerate}

In this standard procedure, we note that as the covariance matrix $\M \Sigma_t$ is calculated globally, no explicit consideration is taken of any potential asymmetries in the return distribution. Our idea is now to utilize the local Gaussian correlation, and one should expect, to be able to improve the total portfolio return. In practice, all steps above are equal, except that in Step 2, the rolling sampling windows of historical returns are used to estimate a local covariance matrix $\M \Sigma_t(\bm x)$ in the gridpoint $\bm x =x_1,...,x_N$, by using the pairwise approach described in the last section. More specifically, step 2 is replaced by the following;

\begin{itemize}
    \item[2.'] During each month $t$, starting from $t=M+1$, returns from the $M$ previous months are used to calculate the one month ahead local covariance matrix $\M \Sigma_t(\bm x)$, consisting of the pairwise local covariances and local standard deviations $\hat{\Sigma}_b(x_i,x_j) = \hat{\rho}_b(x_i,x_j) \hat{\sigma}_{i,b}(x_i) \hat{\sigma}_{j,b}(x_j)$ in the gridpoint $\bm x$ with $b$ being the bandwidth in the local Gaussian approximation. As new information arrives at month $t+1$, we update these estimates. This process is repeated by incorporating the return for each month going forward and ignoring the earliest one, as previously mentioned. The one month ahead expected return vector $\bm \mu_t$ is calculated as in step 2 above (i.e. using the global estimate). 
\end{itemize}

Thus, our model is specified to account for asymmetries by specifying the gridpoint $\bm x$ to use.  But the key question is then; what gridpoint and hence which corresponding local covariance matrix should be used for solving the optimization problem for each time period? In practice, a regular grid is placed across the area of interest, and then an investor can pick any gridpoint based on her preferences. For instance, a risk-averse investor can guard against large losses by selecting a gridpoint representing the asset returns during crisis periods. In this way, the corresponding estimated local covariance matrix will reflect the (historical) dependence structure during crisis periods. However, the selection of the gridpoint can also be dynamic. For instance, the gridpoint may correspond to a subjective meaning of where the investor thinks the market of a particular asset is going to be in the following trading month.
 
In the empirical analysis in section 5, we opt for a simple data-driven selection of gridpoints, where the gridpoint is selected by computing the average of the three last months recorded return observations. Thus the gridpoint will change from one month to the next. More specifically, the "moving-grid" point at time $t$ is defined for all pairs of assets $i,j$ as 

\begin{equation}\label{eq:7.1}
    (x_i,x_j)= \textrm{moving-grid} =(\frac{1}{3}\sum_{k=1}^3 r^i_{t-k}, \frac{1}{3}\sum_{k=1}^3 r^j_{t-k}).
\end{equation}

This is a simple way of letting the covariance matrix dynamically adapt to the dependence structure of the market under the näive assumption that the dependence structure between asset returns in month $t$ corresponds to the dependence structure of asset returns in the neighbourhood of the three months moving average of observed previous returns.   

%In this way we are hopefully able to pick the "most likely" covariance matrix in the following trading month, based on a predicted return of an asset in month $t$ is equal to its three months moving average of observed previous returns. This procedure should produce smoother portfolio weights, than for instance only using the last months observed return as the prediction. But, clearly this should just be taken as a point of departure for future research.  

%%%%%%%%%%%%%%%%%%%%% start empirical analysis %%%%%%%%%%%%%%%%%%%

\section{Data}
Our data set consists of monthly closing prices on six US dollar-denominated indices sourced from Thompson Reuters Datastream. The sample period extends from February 1980 to August 2018, yielding 463 monthly return observations. The included time series are FTSE Actuaries All Share Index, (FTALLSH),  Standard and Poor's 500 Index (S\&P500), UK Benchmark 10 Year DS government bond Index (BMUK10Y), US Benchmark 10 Year DS government bond Index (BMUS10Y), Thomson Reuters Equal Weight Commodity Index (EWCI), and Standard and Poor's GSCI Gold Index (GSGCSPT).

\begin{table}[ht!]
\caption{Overview of the data series}
    \label{tab:data_sets}
    \centering
    \begin{tabular}{ll}
    \toprule
      Name   &  Description \\
     \midrule
     FTALLSH   & FTSE Actuaries All Share Index \\
     S\&P500   & Standard and Poor's 500 Index  \\
     BMUK10Y   & UK Benchmark 10 Year DS government bond Index \\
     BMUS10Y   & US Benchmark 10 Year DS government bond Index \\
     EWCI   & Thomson Reuters Equal Weight Commodity Index \\
     GSGCSPT   & Standard and Poor's GSCI Gold Index \\
     \bottomrule
    \end{tabular}
\end{table}

From the descriptive statistics in Table \ref{tab:stats-and-correlations}, we note that all of the returns are skewed and show relatively high kurtosis. Normality is rejected with the Jarque-Bera test, which is significant on the $1\%$ level for all series. A departure from the Gaussian assumption suggests the multivariate normal distribution with a global covariance matrix may not be a sufficient description of the dependence structure, particularly in the distribution's tails.

The two top panels in Table \ref{tab:stats-and-correlations} show the global and local correlation matrices over the entire sampling period. The latter is constructed for a \textit{bear market} scenario by using the lower $5\%$ percentiles for the grid point selection in the pairwise calculation approach described above. Globally, the strongest positive correlation $\rho=0.76$ can be found between the stock indices FTALLSH and S\&P500, and the strongest negative $\rho=-0.185$ between EWCI and BMUS10Y. Both stock indices show a positive, but close to zero correlation with gold. Locally in the \textrm{bear market} scenario, the positive stock market correlation is larger, $\rho=0.843$, and the negative relation between commodities and US interest rate markets enhanced to $\rho=-0.224$. Here, both stock indices are negatively correlated with gold, with $\rho=-0.135$ and $\rho=-0.131$ for FTALLSH and S\&P500, respectively. Intuitively, this seems reasonable, as gold historically has been considered a safe haven in times of turmoil. Such asymmetries in the returns data will be accounted for by calculating local covariance matrices with the moving-grid approach at each time step for the asset allocation below.

% > apply(ret, 2, function(x) quantile(x, c(0.05)))
%  FTALLSH   S.PCOMP   BMUK10Y   BMUS10Y   NYFECRB   GSGCSPT 
%-7.465709 -7.159218 -3.159496 -3.085890 -5.295045 -7.205235 

\begin{table}[h!]
\centering
\small
\caption{Correlations and descriptive statistics}
\label{tab:stats-and-correlations}
\begin{tabular}{lrrrrrr}
\toprule
 & FTALLSH & S\&P500 & BMUK10Y & BMUS10Y & EWCI & GSGCSPT \\
 \midrule 
 \multicolumn{7}{l}{\textit{Global correlation matrix}} \\ \\
FTALLSH & $1$ & $ $ & $ $  & $ $ & $ $ & $ $ \\ 
S\&P500 & $0.760$ & $1$ & $ $ & $ $ & $ $ & $ $ \\ 
BMUK10Y & $0.184$ & $0.017$ & $1$ & $ $ & $ $ & $ $ \\ 
BMUS10Y & $ $-$0.067$ & $ $-$0.029$ & $0.489$ & $1$ & $ $ & $ $ \\ 
EWCI & $0.246$ & $0.288$ & $ $-$0.094$ & $ $-$0.185$ & $1$ & $ $ \\ 
GSGCSPT & $0.038$ & $0.031$ & $0.080$ & $0.077$ & $0.483$ & $1$ \\ 
\midrule 
\multicolumn{7}{l}{\textit{Local correlation matrix, bear market (lower 5\% percentiles)}} \\ \\
FTALLSH & $1$ & $ $ & $ $ & $ $ & $ $ & $ $ \\ 
S\&P500 & $0.843$ & $1$ & $ $ & $ $ & $ $ & $ $ \\ 
BMUK10Y & $0.174$ & $ $-$0.017$ & $1$ & $ $ & $ $ & $ $ \\ 
BMUS10Y & $0.020$ & $0.034$ & $0.635$ & $1$ & $ $ & $ $ \\ 
EWCI & $0.161$ & $0.185$ & $ $-$0.140$ & $ $-$0.224$ & $1$ & $ $ \\ 
GSGCSPT & $ $-$0.135$ & $ $-$0.131$ & $0.204$ & $0.215$ & $0.480$ & $1$ \\ 
\midrule 
\multicolumn{7}{l}{\textit{Descriptive statistics}} \\ \\
Observations & $463$ & $463$ & $463$ & $463$ & $463$ & $463$ \\ 
Mean & $0.628$ & $0.704$ & $0.769$ & $0.583$ & $0.079$ & $0.177$ \\ 
Std.Dev. & $4.588$ & $4.406$ & $2.376$ & $2.417$ & $3.511$ & $5.211$ \\ 
Variance & $21.050$ & $19.413$ & $5.643$ & $5.839$ & $12.326$ & $27.159$ \\ 
Skewness & $ $-$1.300$ & $ $-$0.968$ & $ $-$0.128$ & $0.453$ & $ $-$0.592$ & $0.026$ \\ 
Kurtosis & $6.288$ & $3.665$ & $1.325$ & $1.960$ & $3.775$ & $3.036$ \\ 
Jarque-Bera & $903.903$ & $335.969$ & $36.135$ & $91.622$ & $306.377$ & $180.971$ \\ 
Sharpe ratio & $0.137$ & $0.160$ & $0.324$ & $0.241$ & $0.023$ & $0.034$ \\ 
Max. drawdown & $49.887$ & $59.811$ & $15.764$ & $12.035$ & $48.397$ & $73.680$ \\ 
Min & $ $-$32.711$ & $ $-$24.677$ & $ $-$7.824$ & $ $-$7.600$ & $ $-$20.050$ & $ $-$21.887$ \\ 
1 Quartile & $ $-$1.474$ & $ $-$1.694$ & $ $-$0.585$ & $ $-$0.922$ & $ $-$1.794$ & $ $-$2.668$ \\ 
Median & $1.176$ & $1.242$ & $0.843$ & $0.497$ & $0.151$ & $ $-$0.161$ \\ 
3 Quartile & $3.559$ & $3.265$ & $2.151$ & $1.853$ & $1.998$ & $2.899$ \\ 
Max & $12.523$ & $14.612$ & $8.851$ & $12.660$ & $13.384$ & $26.336$ \\ 
\bottomrule
\end{tabular}
\end{table}

\section{Empirical results}
Our analysis \footnote{Reproduce results or perform new studies with: \url{https://gitlab.com/sleire/lgportf}} compares the portfolio performance for all MV strategies listed in Table \ref{tab:port_strat} by evaluating outcomes when the optimization is performed with (a) the global covariance matrix and (b) the local covariance matrix calculated with the moving-grid approach. The naive $1/N$ weighted portfolio strategy is used as the benchmark model in the analysis, and we perform the study with sampling windows of $M=120$ and $M=240$ months.

The $1/N$ model distributes weights equally across the portfolio at the start of the sampling period, and is left unadjusted for the rest of the investment horizon. The MVS strategy is the classic approach where historical mean returns and the covariance matrix are used to determine the weights for each out-of-sample period, where no consideration is given within the optimization rule to adjust for estimation error in any form. MVSC is the constrained version, where only positive weights are allowed. The MIN strategy aims to minimize portfolio risk measured as variance of portfolio returns. Finally, MINC is the constrained version, where only positive weights are allowed. All strategies allowing short sales have a lower limit on portfolio weights equal to $-50\%$.

\begin{table}[h!]
    \centering
    \caption{Portfolio strategies}
    \label{tab:port_strat}
    \begin{tabular}{ll}
    \toprule
        Strategy & Description \\
        \midrule
        EW & $1/N$ without rebalancing \\
        MVS & Mean-variance with short sales \\
        MVSC & Mean-variance with short sales constraint \\
        MIN & Minimum variance \\
        MINC & Minimum variance with short sales constraint \\
        \bottomrule
    \end{tabular}
\end{table}

When implemented with the local covariance matrix, the strategies are presented as MVS-L, MVSC-L, MIN-L, MINC-L. The local covariance matrices have been constructed by pairwise correlations with the moving-grid approach, where a simple moving average of length $3$ is used to predict gridpoints for the next month. In the event that the resulting covariance matrix is not positive definite, it is adjusted with the method described in \cite{higham2002computing}. %($t+1$), and we use these forecast as the input for calculating the local Gaussian covariance matrix. 

Inspired by the procedure in \cite{low2013}, we continue with a descriptive analysis of out-of-sample results, followed by an evaluation of portfolio rebalancing, terminal wealth and risk-adjusted performance for each of the strategies.

\subsection{Descriptive statistics portfolio strategies}
Descriptive statistics of the portfolio strategies out-of-sample returns are shown in Table \ref{tab:strategies_descriptive}. We report mean, standard deviation, skewness, kurtosis, minimum value, maximum value, and the maximum portfolio drawdown, which is the
maximum observed loss from a peak to a trough of the portfolio, before a new
peak is attained, for window size $M=120$ (top) and $M=240$ (bottom). 

The mean return tends to increase with the different local Gaussian approaches, and all portfolios achieve moderately higher average returns. In the $M=120$ case, MVS-L reaches the highest mean, followed by MVSC-L, MIN-L, MINC-L, MVS, MVSC, MIN, MVSC and EW. For the $M=240$ window size, the average return ranking is MIN-L, MINC-L, MVS-L, MVSC-L/MINC, MIN, MVS, MVSC and EW. These findings indicate that the local Gaussian approach may be able to capture asymmetries and outperform the corresponding benchmark models.

The lowest standard deviation for $M=120$ is achieved by MINC-L. This is however an exception, as all other strategies have slightly higher values when the local Gaussian method is applied. For $M=240$, all local Gaussian portfolios have moderately higher standard deviations, with the exception of MINC-L. As noted in \cite{low2013}, this can be due to a larger upside variation, which is desirable for investors. We will follow their approach and include downside risk measures when evaluating performance below.

\begin{table}[!htbp] \centering 
\small
\caption{Descriptive statistics portfolio strategies} 
\label{tab:strategies_descriptive} 
\begin{tabular}{lccccccc} 
\toprule
  & Mean & Std.dev. & Skewness & Kurtosis & Min & Max & Max. drawdown \\ 
\midrule
\multicolumn{8}{l}{\textit{Window size M = 120}} \\ \\
EW & $0.423$ & $1.999$ & $ $-$0.714$ & $4.052$ & $ $-$11.916$ & $7.342$ & $22.857$ \\ 
MVS & $0.455$ & $1.492$ & $ $-$0.209$ & $1.645$ & $ $-$5.451$ & $5.778$ & $9.479$ \\ 
MVSC & $0.444$ & $1.486$ & $ $-$0.256$ & $1.644$ & $ $-$5.451$ & $5.688$ & $9.479$ \\ 
MIN & $0.435$ & $1.426$ & $ $-$0.190$ & $1.211$ & $ $-$5.108$ & $5.382$ & $9.404$ \\ 
MINC & $0.427$ & $1.430$ & $ $-$0.195$ & $1.181$ & $ $-$5.108$ & $5.382$ & $9.404$ \\ 
MVS-L & $0.491$ & $1.539$ & $ $-$0.214$ & $1.693$ & $ $-$5.701$ & $6.177$ & $8.959$ \\ 
MVSC-L & $0.484$ & $1.494$ & $ $-$0.156$ & $1.592$ & $ $-$5.020$ & $6.079$ & $8.881$ \\ 
MIN-L & $0.462$ & $1.459$ & $ $-$0.218$ & $1.182$ & $ $-$5.194$ & $5.429$ & $8.577$ \\ 
MINC-L & $0.460$ & $1.401$ & $ $-$0.080$ & $1.222$ & $ $-$4.620$ & $5.743$ & $8.065$ \\ 
\midrule
\multicolumn{8}{l}{\textit{Window size M = 240}} \\ \\
EW & $0.376$ & $2.158$ & $ $-$0.750$ & $4.356$ & $ $-$11.916$ & $7.342$ & $22.857$ \\ 
MVS & $0.447$ & $1.671$ & $ $-$0.672$ & $3.012$ & $ $-$8.294$ & $5.061$ & $15.677$ \\ 
MVSC & $0.440$ & $1.670$ & $ $-$0.703$ & $3.047$ & $ $-$8.294$ & $5.122$ & $15.677$ \\ 
MIN & $0.468$ & $1.619$ & $ $-$0.564$ & $2.316$ & $ $-$7.493$ & $5.053$ & $13.530$ \\ 
MINC & $0.470$ & $1.624$ & $ $-$0.558$ & $2.272$ & $ $-$7.476$ & $5.053$ & $13.524$ \\ 
MVS-L & $0.488$ & $1.765$ & $ $-$0.243$ & $3.461$ & $ $-$8.522$ & $7.516$ & $15.354$ \\ 
MVSC-L & $0.470$ & $1.737$ & $ $-$0.348$ & $3.434$ & $ $-$8.522$ & $7.182$ & $15.364$ \\ 
MIN-L & $0.503$ & $1.673$ & $0.805$ & $7.357$ & $ $-$6.023$ & $11.198$ & $10.458$ \\ 
MINC-L & $0.495$ & $1.604$ & $0.012$ & $1.981$ & $ $-$5.926$ & $7.142$ & $11.036$ \\ 
\bottomrule
\end{tabular} 
\end{table}

All strategy returns exhibit slight negative skewness, except for MIN-L and MINC-L for the $M=240$ window, with values of $0.805$ and $0.012$ respectively. Disregarding the EW strategies, the largest negative skew of $-0.703$ can be found in MVSC, for $M=240$. Hence, the strategies are all moderately skewed, or approximately symmetric.

The MIN-L for $M=240$ holds the largest kurtosis value in the analysis. When examining the minimum and maximum returns for this strategy, we observe larger values for both. The MIN-L also achieves the lowest drawdown for $M=240$. The smallest maximum drawdown for $M=120$ is produced by MINC-L. Overall, the local gaussian strategies all have lower maximum drawdowns when compared to their benchmarks, in both windows.

\subsection{Portfolio rebalancing and terminal wealth}

The primary goal of a rebalancing strategy is to minimize risk relative to the target asset allocation produced by the trading strategy. According to \cite{tokat2007portfolio}, the asset manager needs to consider 1) frequency of rebalancing; 2) how large deviations to accept before triggering rebalancing; and 3) whether to restore a portfolio to its target or to some intermediate allocation. Inability to fully rebalance towards the target portfolio weights will lead to sub-optimal diversification. A decision-maker facing practical limitations such as regulatory requirements and periods with weak market liquidity will find strategies with stable target portfolio weights easier to implement relative to those who require more trading, \cite{demiguel2007}. In our study, portfolios are fully rebalanced to target weights on a monthly basis. We evaluate differences in required trading activity and associated transaction costs.

Table \ref{tab:strategies_rebal_wealth} provides a summary of the portfolio rebalancing analysis and the terminal wealth reached by each of the strategies. It shows the average standard deviation within target portfolio weights, maximum positive and maximum negative adjustments of weights, average turnover, and terminal wealth of a hypothetical investment of \$ 1 for each strategy. The average standard deviation within target portfolio weights across the entire out-of-sample time period is calculated as follows:

\begin{equation}
    \bar \sigma(\hat{w}_{k, M}) = \frac{\sum_{t=1}^{n-M} \sigma(\hat{w}_{k, t, M})}{n-M},
\end{equation}
where
\begin{equation}
    \sigma(\hat{w}_{k, t, M}) = \sqrt{\frac{1}{N} \sum_{i=1}^{N} (\hat{w}_{k, t, M, i} - \bar{w}_{k, t, M})^2}
\end{equation}
where 
%$\sigma(\hat{w}_{k, t, M})$ is the $N$ vector of portfolio weights at time $t$ for strategy $k$ based upon a window sampling size of $M$ months. Furthermore, 
$\hat{w}_{k, t, M, i}$ is is the portfolio weight for asset $i$ in a portfolio of $N$ assets for strategy $k$ based upon a window sampling of $M$ months, and $\bar{w}_{k, t, M}$ is the average portfolio weight across the $N$ assets in the portfolio. The maximum values for positive and negative weight adjustments are selected by identifying the largest positive and negative weight changes on the asset level. Following \cite{demiguel2007}, we also report the average turnover, which is calculated  as the average sum of the absolute value of the transactions over the $N$ assets with:

\begin{equation}
    \textrm{Average turnover} = \frac{1}{n-M} \sum_{t=1}^{n-M} \sum_{j=1}^{N} (|w_{k,j,t+1}-w_{k,j,t⁺}|),
\end{equation}
where $N$ is the number of assets in the portfolio, $n$ is the full length of the returns series, $M$ is the window size, $w_{k,j,t+1}$ is the target weight for asset $j$ at time $t+1$ for strategy $k$, and $w_{k,j,t⁺}$ is the corresponding asset weight before rebalancing. Terminal wealth is calculated assuming no transaction costs, and with a transaction cost of $1$ basis point.

\begin{table}[!h] \centering 
\small
\caption{Portfolio rebalancing and terminal wealth} 
\label{tab:strategies_rebal_wealth} 
\begin{tabular}{lcccccc} 
\toprule
  & $\bar \sigma(\hat{w}_{k, M})$ & Max. adj. & Min. adj. & Avg.turnover &  Wealth & Wealth incl.tcost\\ 
\midrule
\multicolumn{7}{l}{\textit{Window size M = 120}} \\ \\
EW & $0$ & $0$ & $0$ & $0$ & $4.052$ & $4.052$ \\ 
MVS & $20.671$ & $18.759$ & $ $-$14.112$ & $8.402$ & $4.680$ & $4.548$ \\ 
MVSC & $17.944$ & $18.737$ & $ $-$17.349$ & $7.117$ & $4.506$ & $4.399$ \\ 
MIN & $19.332$ & $6.966$ & $ $-$8.769$ & $4.135$ & $4.376$ & $4.315$ \\ 
MINC & $18.536$ & $6.966$ & $ $-$8.769$ & $3.533$ & $4.254$ & $4.203$ \\ 
MVS-L & $20.760$ & $122.681$ & $ $-$132.874$ & $32.084$ & $5.283$ & $4.736$ \\ 
MVSC-L & $16.521$ & $29.529$ & $ $-$30.509$ & $17.045$ & $5.166$ & $4.875$ \\ 
MIN-L & $21.174$ & $117.820$ & $ $-$147.244$ & $38.222$ & $4.775$ & $4.192$ \\ 
MINC-L & $17.710$ & $45.461$ & $ $-$46.261$ & $18.301$ & $4.747$ & $4.460$ \\ 
\midrule
\multicolumn{7}{l}{\textit{Window size M = 240}} \\ \\
EW & $0$ & $0$ & $0$ & $0$ & $2.231$ & $2.231$ \\ 
MVS & $16.663$ & $9.228$ & $ $-$6.741$ & $6.325$ & $2.605$ & $2.569$ \\ 
MVSC & $15.813$ & $9.228$ & $ $-$7.426$ & $5.622$ & $2.571$ & $2.539$ \\ 
MIN & $16.249$ & $5.036$ & $ $-$4.434$ & $3.090$ & $2.729$ & $2.711$ \\ 
MINC & $15.952$ & $5.039$ & $ $-$4.428$ & $2.793$ & $2.742$ & $2.725$ \\ 
MVS-L & $15.660$ & $73.184$ & $ $-$89.316$ & $16.967$ & $2.855$ & $2.750$ \\ 
MVSC-L & $14.253$ & $18.096$ & $ $-$26.245$ & $12.105$ & $2.745$ & $2.672$ \\ 
MIN-L & $17.014$ & $53.636$ & $ $-$79.931$ & $21.617$ & $2.953$ & $2.815$ \\ 
MINC-L & $15.243$ & $19.646$ & $ $-$26.028$ & $13.762$ & $2.910$ & $2.823$ \\ 
\bottomrule
\end{tabular} 
\end{table}

The variability of portfolio weights reported in Table \ref{tab:strategies_rebal_wealth} shows the larges values for the MIN-L strategy, both for $M=120$ and $M=240$. Results for the remaining strategies are mixed. The local Gaussian models does not seem to systematically achieve either higher or lower average standard deviation in target portfolio weights compared to their benchmarks. Looking at the maximum and minimum adjustments of portfolio weights however, there are clear differences. The local Gaussian strategies require adjustments of larger magnitude, in both directions. This is particularly the case for the unconstrained models allowing short sales. For example, during a period of large market moves, the MIN-L strategy exploit nearly it's full mandate with a maximum negative weight adjustment of $-147\%$ for one of the assets in the $120$ window. This is a significant adjustment, that does generate additional costs. Viewed across all strategies, we see larger and more frequent adjustments, resulting in an average turnover $2.7-9.2$ times higher than the classical MV portfolios when the lowest negative weight allowed is set to $-50\%$. For the long only portfolios, the differences are smaller, but still significant. Average turnover is increased by a factor of $2.2-5.2$.

The increase in traded volume translates into lower terminal wealth when transaction costs are included in the analysis. While all local Gaussian strategies achieve larger terminal wealth when disregarding costs of trade, the MIN-L for $M=120$ shows weakest performance when 1 basis point is added as a transaction fee. The remaining local Gaussian strategies still reach a larger terminal wealth. Top-ranked strategies ex. costs are MVS-L ($M=120$) and MIN-L ($M=240$). When costs are included, these are replaced by the long only portfolios MVSC-L and MINC-L. These two achieve a final wealth which of $10.8\%$ and $10.4\%$ larger than their classical MV benchmarks.

\newpage

\begin{figure}[!h]  
\begin{minipage}{0.85\textwidth} 
\caption{Wealth accumulation $M=120$}   
\includegraphics[width=\linewidth]{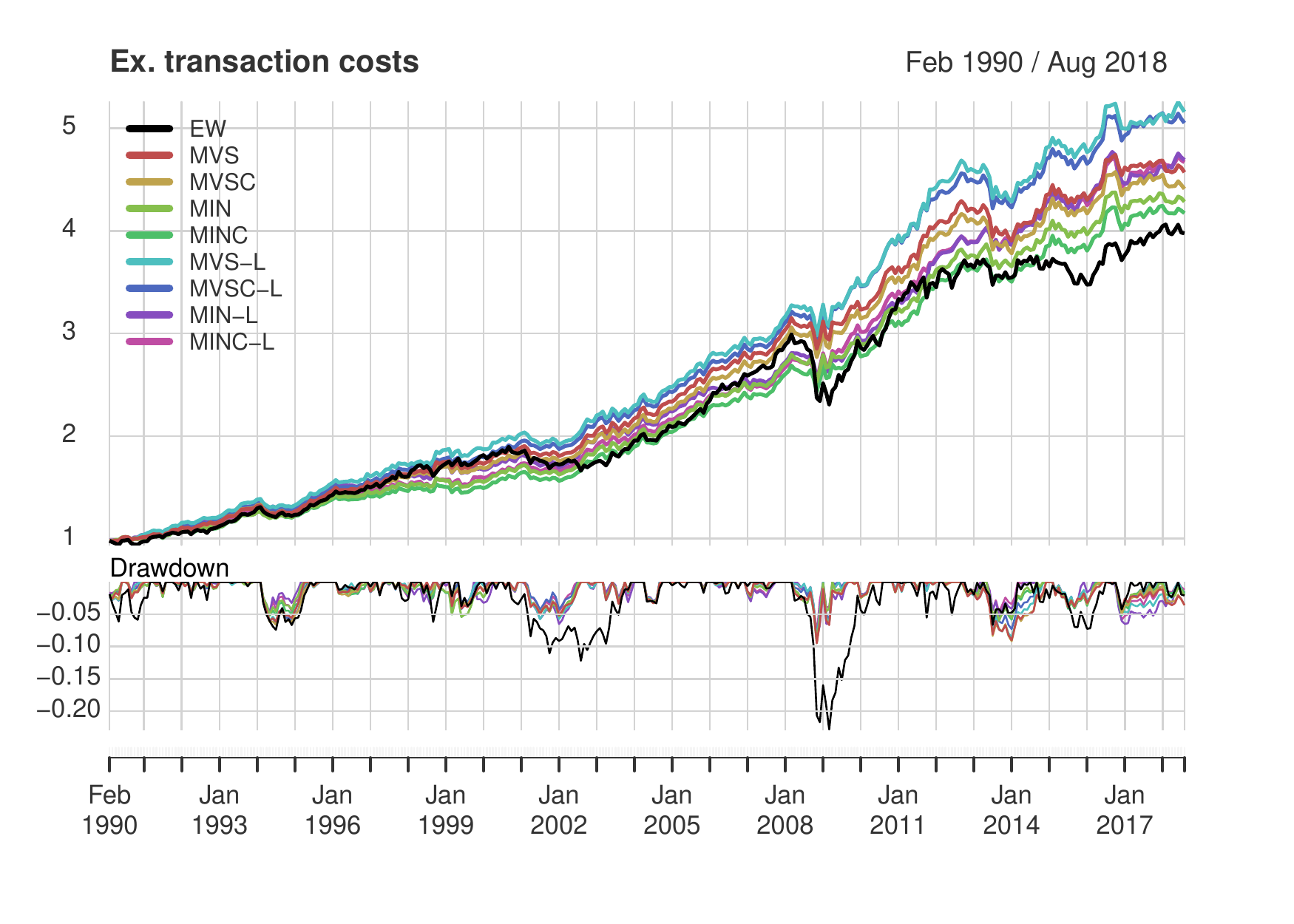}  
\label{fig:wealth_drawdown_120}
\end{minipage}    
\begin{minipage}{0.85\textwidth} 
\includegraphics[width=\linewidth]{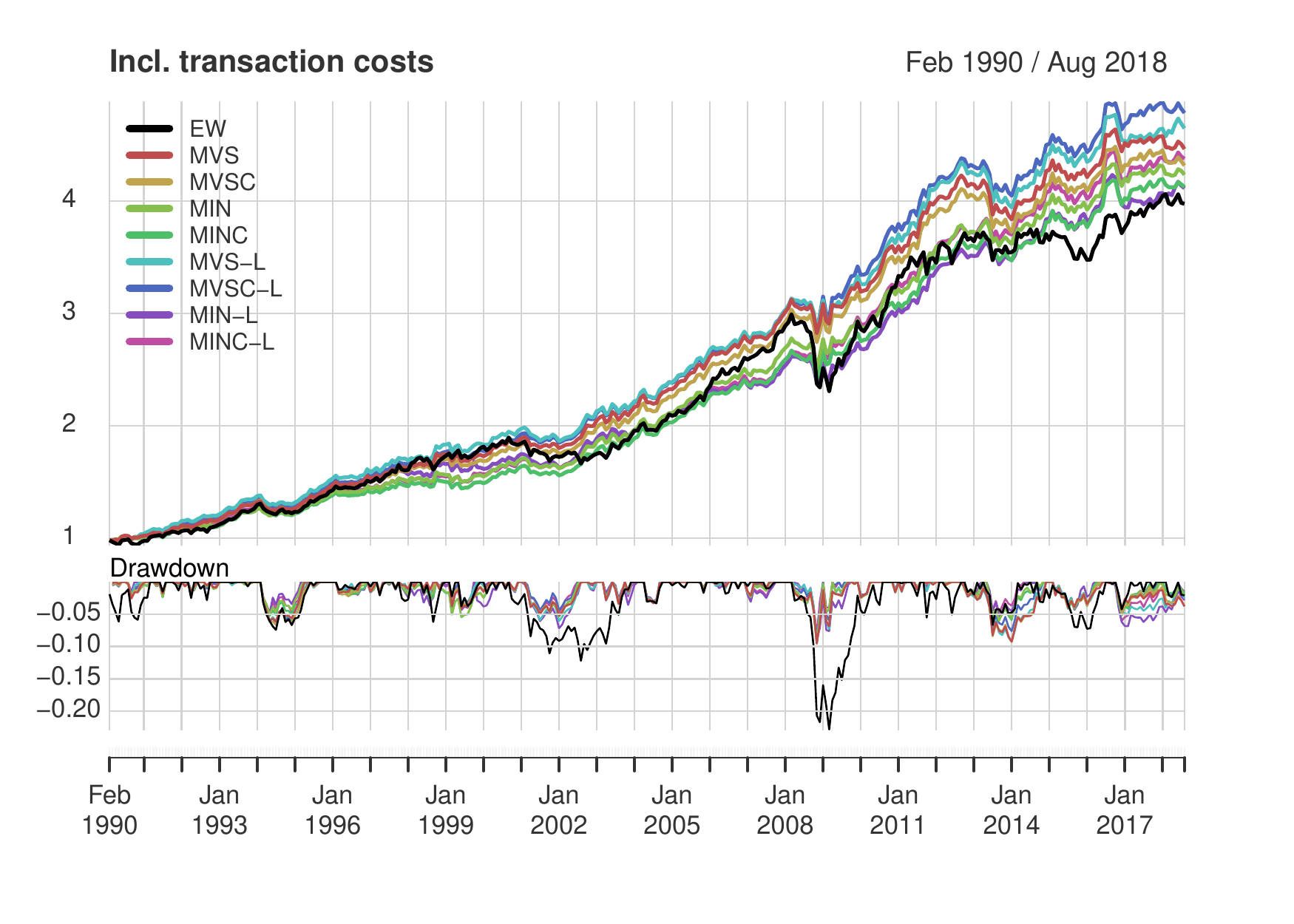}    
\end{minipage}   
\end{figure}

\begin{figure}[!h]  
\begin{minipage}{0.85\textwidth} 
\caption{Wealth accumulation $M=240$}   
\includegraphics[width=\linewidth]{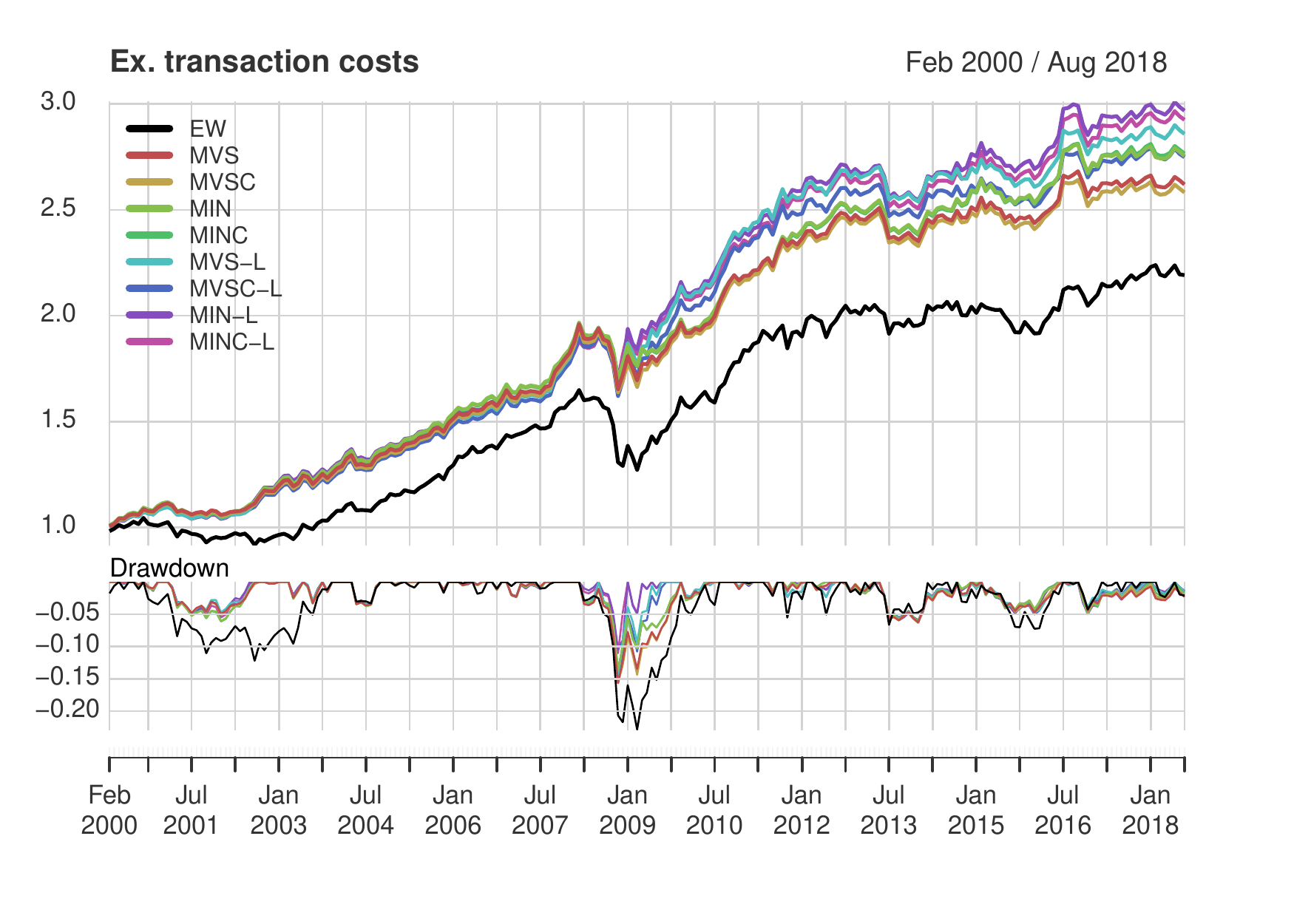}  
\label{fig:wealth_drawdown_240}
\end{minipage}    
\begin{minipage}{0.85\textwidth} 
\includegraphics[width=\linewidth]{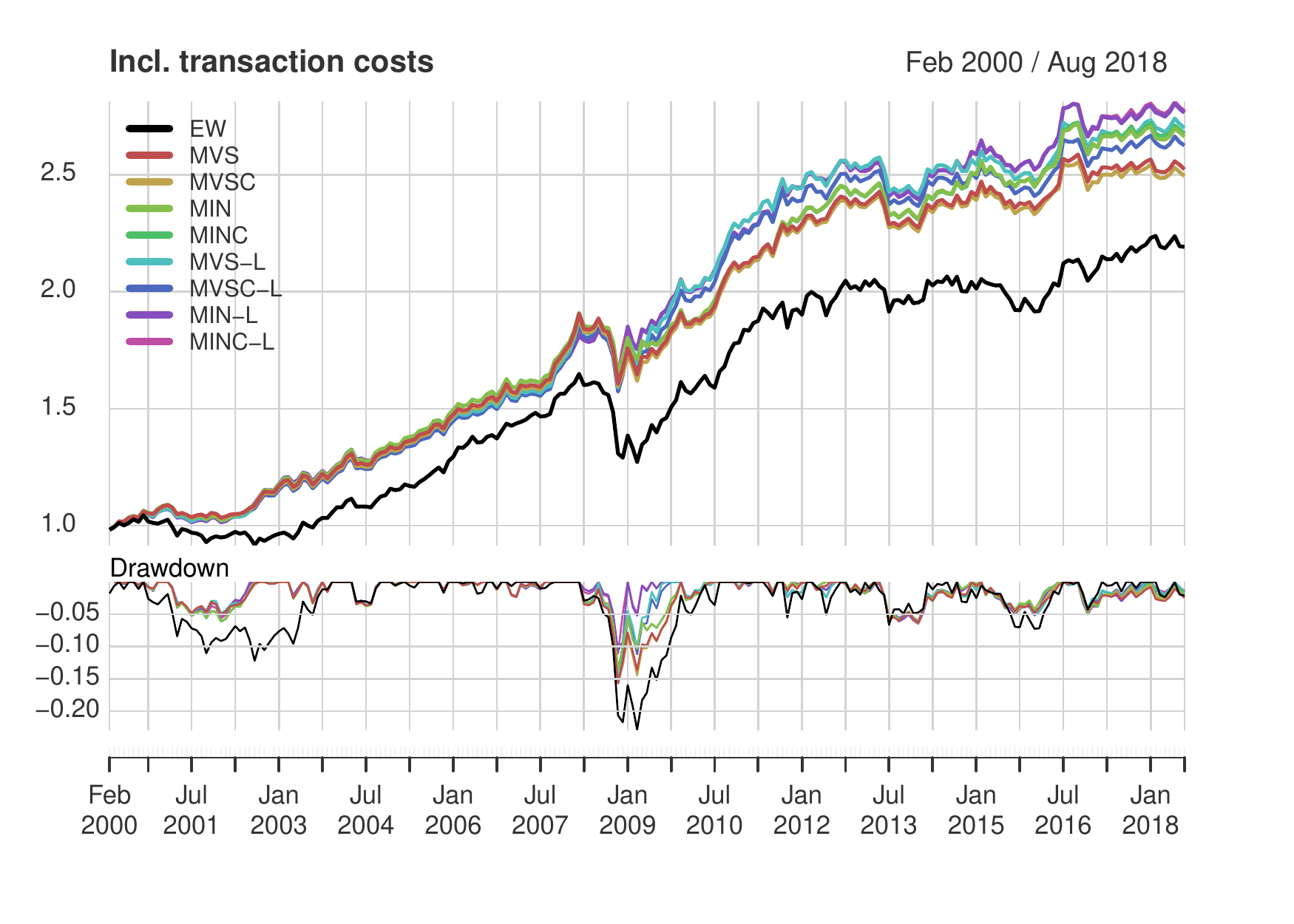}    
\end{minipage}   
\end{figure}

\newpage

Figure \ref{fig:wealth_drawdown_120} shows wealth accumulation and drawdowns for the hypothetical investment of \$ 1 in each of the nine strategies included in the analysis when a sampling window of $M=120$ is used. As seen in the upper part of the figure, the local Gaussian MVS-L produces the largest final wealth when disregarding costs of trade. It remains top-ranked during most months in the sample, and suffers from smaller drawdowns in volatile periods such as the 2008 financial crisis. When transaction costs are considered, the strategy still performs well, but is pushed down from the top position by the constrained MVSC-L, which has lower turnover.
A similar illustration for $M=240$ can be found in Figure \ref{fig:wealth_drawdown_240}.  

\subsection{Evaluation of risk-adjusted performance}

Table \ref{tab:strategies_performance_incl_tcost} reports out-of-sample performance by evaluating portfolio strategy returns using a range of risk-adjusted metrics. The traditional Sharpe ratio was introduced as a measure for mutual fund performance in \cite{sharpe1966mutual} under the term \textit{reward-to-variability ratio}. The ratio is a risk-adjusted measure of return using standard deviation to represent risk. As standard deviation also will penalize upside deviations in returns, several modifications have been suggested. In VaR Sharpe and ES Sharpe, the Value at Risk and Expected Shortfall are used as risk measures. The Certainty Equivalent (CEQ) assuming quadratic utility with a risk aversion parameter equal to $\gamma = 1$ is also included. The Sortino ratio introduced in  \cite{sortino1994performance} penalize downside standard deviation only. Finally, the Omega ratio from \cite{keating2002universal} is calculated as a probability-weighted ratio of gains versus losses for a threshold return target (here set to zero), without making any assumptions regarding investors utility or risk aversion. All metrics produce high values for the best-performing strategies.

Results excluding transaction costs are reported in Panel A. The classic Sharpe ratio using standard deviation as risk measure is higher for all portfolios with the local Gaussian approach, and the largest value is attained by the MINC-L in both windows. This is also reflected in the annualized Sharpe ratio. For windows $M=120$ and $M=240$, the other top rankings are: MINC-L and MIN-L (VaR Sharpe), MINC-L and MIN-L (ES Sharpe ratio), MVS-L and MIN-L (CEQ), MINC-L and MIN-L (Sortino ratio), MINC-L and MIN-L (Omega ratio). With the exception of MVS-L, all the top-ranked strategies are of type \textit{minimum-variance}, optimized with the local covariance matrix. In a total of 14 top rankings, 8 are held by strategies with a long-only constraint. For $M=120$, 6 out of 7 are long-only, while for $M=240$, 2 out of 7 does not allow short sales.

Results including transaction costs are reported in Panel B. Here, the MINC-L still produces the largest Sharpe ratio in both windows. Improvements in Sharpe ratios are reduced for some of the unconstrained strategies in the $M=120$ window. The MVS-L and MIN-L do not manage to achieve larger scores than their benchmarks evaluated with Sharpe, VaR Sharpe, ES Sharpe, Sortino and Omega ratios. In the $M=240$ window, all local Gaussian models still perform slightly better than their benchmarks.

\begin{table}[htbp!] \centering 
%\small
\caption{Portfolio strategies out-of-sample performance} 
\label{tab:strategies_performance_incl_tcost} 
%\scalebox{0.7}{
\resizebox{0.82\columnwidth}{!}{%
\begin{tabular}{lccccccc} 
\toprule
   & Sharpe & VaR Sharpe & ES Sharpe & Ann. Sharpe & CEQ & Sortino & Omega \\ 
\midrule
%\hline \\[-1.8ex] 
\multicolumn{8}{l}{\textbf{Panel A: Ex. transaction costs}} \\ \\
\midrule
\multicolumn{8}{l}{\textit{Window size M = 120}} \\ \\
EW & $0.212$ & $0.137$ & $0.071$ & $0.715$ & $0.403$ & $0.327$ & $1.737$ \\ 
MVS & $0.305$ & $0.224$ & $0.143$ & $1.057$ & $0.444$ & $0.525$ & $2.221$ \\ 
MVSC & $0.299$ & $0.216$ & $0.138$ & $1.035$ & $0.433$ & $0.508$ & $2.191$ \\ 
MIN & $0.305$ & $0.223$ & $0.148$ & $1.057$ & $0.425$ & $0.528$ & $2.199$ \\ 
MINC & $0.299$ & $0.217$ & $0.145$ & $1.034$ & $0.417$ & $0.514$ & $2.169$ \\ 
MVS-L & $0.319$ & $0.237$ & $0.150$ & $1.108$ & $\textbf{0.479}$ & $0.557$ & $2.298$ \\ 
MVSC-L & $0.324$ & $0.244$ & $0.157$ & $1.126$ & $0.473$ & $0.569$ & $2.315$ \\ 
MIN-L & $0.317$ & $0.232$ & $0.154$ & $1.099$ & $0.451$ & $0.555$ & $2.264$ \\ 
MINC-L & $\textbf{0.328}$ & $\textbf{0.250}$ & $\textbf{0.168}$ & $\textbf{1.141}$ & $0.450$ & $\textbf{0.587}$ & $\textbf{2.340}$ \\ 
\midrule
\multicolumn{8}{l}{\textit{Window size M = 240}} \\ \\
EW & $0.174$ & $0.110$ & $0.057$ & $0.577$ & $0.353$ & $0.263$ & $1.593$ \\ 
MVS & $0.267$ & $0.179$ & $0.099$ & $0.919$ & $0.433$ & $0.427$ & $2.025$ \\ 
MVSC & $0.264$ & $0.175$ & $0.097$ & $0.905$ & $0.426$ & $0.417$ & $2.009$ \\ 
MIN & $0.289$ & $0.198$ & $0.116$ & $0.999$ & $0.455$ & $0.474$ & $2.120$ \\ 
MINC & $0.290$ & $0.199$ & $0.117$ & $1.000$ & $0.457$ & $0.474$ & $2.120$ \\ 
MVS-L & $0.276$ & $0.203$ & $0.115$ & $0.951$ & $0.472$ & $0.472$ & $2.126$ \\ 
MVSC-L & $0.270$ & $0.193$ & $0.107$ & $0.929$ & $0.454$ & $0.453$ & $2.075$ \\ 
MIN-L & $0.301$ & $\textbf{0.317}$ & $\textbf{0.317}$ & $1.041$ & $\textbf{0.489}$ & $\textbf{0.567}$ & $\textbf{2.303}$ \\ 
MINC-L & $\textbf{0.309}$ & $0.240$ & $0.157$ & $\textbf{1.070}$ & $0.482$ & $0.554$ & $2.268$ \\ 
\midrule
%\hline \\[-1.8ex] 
\multicolumn{8}{l}{\textbf{Panel B: Incl. transaction costs}} \\ \\
\midrule
\multicolumn{8}{l}{\textit{Window size M = 120}} \\ \\
EW & $0.215$ & $0.140$ & $0.072$ & $0.727$ & $0.410$ & $0.332$ & $1.753$ \\ 
MVS & $0.306$ & $0.225$ & $0.143$ & $1.061$ & $0.444$ & $0.527$ & $2.230$ \\ 
MVSC & $0.300$ & $0.218$ & $0.138$ & $1.040$ & $0.434$ & $0.511$ & $2.202$ \\ 
MIN & $0.309$ & $0.226$ & $0.150$ & $1.070$ & $0.428$ & $0.535$ & $2.221$ \\ 
MINC & $0.302$ & $0.220$ & $0.146$ & $1.046$ & $0.421$ & $0.520$ & $2.190$ \\ 
MVS-L & $0.304$ & $0.222$ & $0.139$ & $1.054$ & $0.456$ & $0.523$ & $2.213$ \\ 
MVSC-L & $0.319$ & $0.239$ & $0.153$ & $1.108$ & $\textbf{0.464}$ & $0.558$ & $2.289$ \\ 
MIN-L & $0.295$ & $0.211$ & $0.139$ & $1.020$ & $0.420$ & $0.505$ & $2.140$ \\ 
MINC-L & $\textbf{0.320}$ & $\textbf{0.242}$ & $\textbf{0.162}$ & $\textbf{1.111}$ & $0.438$ & $\textbf{0.568}$ & $\textbf{2.295}$ \\ 
\midrule
\multicolumn{8}{l}{\textit{Window size M = 240}} \\ \\
EW & $0.179$ & $0.113$ & $0.058$ & $0.593$ & $0.362$ & $0.270$ & $1.613$ \\ 
MVS & $0.263$ & $0.175$ & $0.097$ & $0.902$ & $0.426$ & $0.419$ & $1.999$ \\ 
MVSC & $0.260$ & $0.172$ & $0.095$ & $0.891$ & $0.421$ & $0.410$ & $1.987$ \\ 
MIN & $0.286$ & $0.195$ & $0.115$ & $0.985$ & $0.450$ & $0.467$ & $2.100$ \\ 
MINC & $0.286$ & $0.196$ & $0.116$ & $0.988$ & $0.453$ & $0.468$ & $2.102$ \\ 
MVS-L & $0.267$ & $0.193$ & $0.109$ & $0.917$ & $0.457$ & $0.451$ & $2.069$ \\ 
MVSC-L & $0.263$ & $0.187$ & $0.103$ & $0.904$ & $0.444$ & $0.438$ & $2.034$ \\ 
MIN-L & $0.287$ & $\textbf{0.291}$ & $\textbf{0.291}$ & $0.992$ & $0.467$ & $\textbf{0.534}$ & $\textbf{2.214}$ \\ 
MINC-L & $\textbf{0.299}$ & $0.230$ & $0.151$ & $\textbf{1.035}$ & $\textbf{0.468}$ & $0.532$ & $2.208$ \\ 
\bottomrule
%\hline \\[-1.8ex] 
%\multicolumn{8}{l}{Risk-adjusted performance metrics, maximum values in \textbf{bold}}
\end{tabular} 
}
\end{table}
For windows $M=120$ and $M=240$, the remaining top rankings are: MINC-L and MIN-L (VaR Sharpe), MINC-L and MIN-L (ES Sharpe ratio), MVSC-L and MINC-L (CEQ), MINC-L and MIN-L (Sortino ratio), MINC-L and MIN-L (Omega ratio).

In this analysis, the local Gaussian approach seems to improve performance. This suggests that challenges related to return asymmetries may be handled in a familiar and well-established framework for portfolio management by replacing the global covariance matrix with it's a local cousin. Improved performance and simplicity is some of the appeal with the local Gaussian approach to portfolio management in the MV setting. There are however matters to keep in mind when implementing the approach. The selection of gridpoints for calculating the pairwise local correlations will affect the local Gaussian covariance matrix. We have evaluated alternative approaches to the moving gridpoint selection without observing substantial changes in results and conclusions. A more thorough analysis of these effects is left for future studies.

\section{Summary and conclusion}

In this paper, we investigate whether the asymmetries typically found in financial returns data can be modeled using a new non-parametric  measure  of  local  dependence to improve asset allocation in a traditional Mean-Variance portfolio setting. \cite{markowitz1952} explicitly recommends the use of a probability model to generate the model inputs. Our study focuses on improving the covariance matrix used as input to a range of MV optimization rules by utilizing a model-based approach, namely the local Gaussian correlation.

We investigate the performance of the $1/N$ benchmark and four MV portfolio strategies in a six-asset portfolio consisting of indices exposed to stocks, commodities and interest rates. The strategies are implemented both with the global covariance matrix and the corresponding local covariance matrix calculated with a moving-grid method. The analysis is performed on monthly returns data with sampling windows of $120$ and $240$ observations. The strategies are evaluated out-of-sample based on need for portfolio rebalancing, turnover, terminal wealth and risk-adjusted performance.

Our findings suggest that the local Gaussian approach to portfolio management improves the traditional MV portfolio optimization when asymmetries are present in asset returns data. When investigating portfolio rebalancing, we find that the local Gaussian strategies do require higher turnover. After transactions costs have been taken into account, seven out of eight strategies achieve higher terminal wealth relative to their MV-benchmarks. In the risk-adjusted performance evaluation disregarding transaction costs, all local Gaussian strategies outperform the corresponding MV-models. When the cost of trade is included, six out of eight strategies improve upon their benchmarks, and the top-performing strategies are of local Gaussian type. This suggests the proposed methodology may be a viable and straightforward approach for improving asset allocation when asymmetries are present in returns data.

%when evaluated with Sharpe ratio, VaR Sharpe ratio, ES Sharpe ratio, the Certainty Equivalent with quadratic utility, Sortino ratio and Omega ratio. 

\section*{Acknowledgements}  
This work has been partly supported by the Finance Market Fund (Norway). 
%We thank a referee and the Editors for suggestions leading to an improved presentation.

%\newpage

\singlespacing
\bibliographystyle{plainnat}
\bibliography{bibunblind}

\end{document}